**RESEARCH**

# A New Alternative for Traffic Hotspot Localization in Wireless Networks Using O&M Metrics

Jaziri Aymen[1,2*], Nasri Ridha[1] and Chahed Tijani[2]

**Abstract**

In recent years, there has been an increasing awareness to traffic localization techniques driven by the problematic of hotspot offloading solutions, the emergence of heterogeneous networks (HetNet) with small cells' deployment and the green networks. The localization of traffic hotspots with a high accuracy is indeed of great interest to know how the congested zones can be offloaded, where small cells should be deployed and how they can be managed for sleep mode concept. We propose, in this paper, a new hotspot localization technique based on the direct exploitation of five Key Performance Indicators (KPIs) extracted from the Operation and Maintenance (O&M) database of the network. These KPIs are the Timing Advance (TA), the angle of arrival (AoA), the neighboring cell level, the load time and two mean throughputs: arithmetic (AMT) and harmonic (HMT). The combined use of these KPIs, projected over a coverage map, yields a promising localization precision and can be further optimized by exploiting commercial data on potential hotspots. This solution can be implemented in the network at an appreciable low cost when compared with widely used probing methods.

*Correspondence: aymen.jaziri@orange.com
[1]Orange Labs, 38/40 avenue General Leclerc, 92794 Issy-les-Moulineaux, France
[2]Telecom SudParis, 9 street Charles Fourier, 91011 Evry, France
Full list of author information is available at the end of the article

## 1 Introduction

Localization of traffic hotspots is often one of the first steps in network planning and optimization, especially in the context of newly proposed technologies within 3GPP standards such as Heterogeneous Networks (HetNets) [2]. HetNets are composed of small cells which are to be deployed, in addition to macro cells, in areas of capacity bottlenecks representing typical hotspots. And so, the efficiency of the deployed solution to absorb traffic in the congested zone obviously depends on the accuracy of the localization of the traffic hotspot zones. Several traffic localization techniques have been proposed in 3GPP Long Term Evolution (LTE) systems. They are mostly based on probing and include trace analysis and protocol decoding [3–5]. The most important information extracted from traces is the received power level and the timing advance. Authors in [3] provided a test transmitter which plays the role of a neighboring cell at first and then the role of a serving Base Station (BS). Tests are realized within the existing cells in the network in order to assess the traffic density within the vicinity of the transmitter means. This solution needs to configure the test transmitter and realize measurements in each area separately. This method may take a long time to assess the traffic distribution in the entire zone covered by a BS. In the same context, patent [4] disclosed a method where the User Equipments (UEs) send periodically a report of radio measurements from the serving cell and the neighboring cells. A recording unit post is installed at the interface between the BS and the studied Base Station Controller (BSC), termed A-bis, to examine the messages exchanged on this interface. Based on these measurements, the traffic distribution is calculated. It was shown that the precision using this method does not fit with the small cell dimensions. Another alternative method was presented in patent [5] and treated traffic hotspot localization in GSM using statistical analysis of the timing advance and neighboring cell measurements extracted from traces. The accuracy of this method is improved as compared to the previous ones but it is still insufficient because it only localizes the number of UEs and omit the data volume. Furthermore, proximity location is another method of localization based on the detection of close Wi-Fi Access Points (APs). This helps to calculate the position of the UE knowing that of the AP [6,9]. RF-Fingerprinting [7,9], cell-ID (LTE Rel 8) and A-GPS (LTE Rel 9) [9] are also well known techniques for locating individual UEs with different accuracy levels. However, these individual locating methods are quite complex (in the generation of a spatial traffic distribution) because it involves a large number of UEs, from which information about hotspot traffic distribution is extracted.

In sum, the probing-based methods suffer from several shortcomings: First, not all the traces are captured by probes, some of them are lost. Second, they require high capacity storage servers that can not be provided by existing Operation and Maintenance (O&M) databases. Third, based on recorded traces, the position of each UE is calculated separately leading thus to a heavy localization process. Eventually, the tools needed for probing, storage and processing are costly.

Another family of localization methods makes use of Key Performance Indicators (KPIs) in order to infer the level of traffic in each area of the cell. In [8], authors proposed to divide each cell of the network into several subareas, and to calculate a traffic value for each subarea from O&M measurements in the network. The indicators used to compute traffic values are the load of the cell, the call attempts and the number of handovers. This method is very simple and does not require additional tests or equipment. However, it is limited in high data rate networks, such as 4G systems, because it only localizes traffic hotspots in terms of number of UEs. The idea of [8] can be further improved considering other KPIs. Besides, the traffic localization, in the way described in the paper, is based on the definition of an optimization problem where the number of variables to find is the number of subareas. In this case, the method has a significant computational cost mainly for a precision going to small cell dimensions.

In order to improve the method used in [8] with fast computations, we propose, in this paper, a new method for hotspot localization based on the combined use of several KPIs directly obtained from the O&M. These KPIs are: Timing Advance (TA), Angle of Arrival (AoA), neighboring cell level, load time and two mean throughputs: arithmetic (AMT) and harmonic (HMT). Simulation results show that the proposed method in this paper achieves acceptable localization along with sufficient precision and significant savings on the cost of localization, as compared to probing-based techniques.

The main contributions of this paper are threefold. The first main contribution is the use of AoA, the correlation between the neighboring cells' traffic loads and the difference between the AMT and HMT, in addition to the traditionally used TA and neighboring cell level and the projection of the information extracted from these KPIs over the coverage map. The second novelty of this paper is the definition of a global metric combining all the five, previously-cited KPIs. This metric is further optimized using additional information about potential hotspot zones obtained from commercial data. Third, smoothing the estimated map is an additional step proposed in this paper in order to make the estimated traffic distribution more precise.

The remainder of this paper is organized as follows. A general description of the process of traffic localization is provided and the key motivations behind the use of the KPIs are given in Section II. Then, the three main inputs of the hotspot localization algorithm are detailed in section III. In Section IV, we present the hotspot localization algorithm and its optimization. Section V contains simulations and corresponding results. Sections VI eventually concludes the paper.

## 2 Traffic localization method: Process and key motivations for using KPIs

The proposed algorithm of hotspot localization is depicted in Fig. 1. Five KPIs define the first main inputs of the algorithm: TA, AoA, neighboring cell level, correlation between traffic loads and the difference between the AMT and HMT. Extracting information from these 5 KPIs and projecting them over the coverage map, as will be explained later in Section 4, enables us to obtain the traffic level in each pixel[1] of the coverage map.

The 5 KPIs are properly chosen with respecting two important criteria: locating hotspots in terms of number of connected UEs and traffic volume. In fact, TA, AoA and neighboring cell level can be sufficient for calculating the spatial distribution of connected UEs. However, in practice, a hotspot is measured by introducing not only the high density of connected UEs but also the traffic volume. Actually, this latter does not follow exactly the same evolution as the number of connected UEs. Therefore, the traffic hotspot localization is improved by adding directly data-volume related KPIs such as the correlation between the cells' traffic load and the difference between the AMT and HMT.

---

[1]The coverage map of the network is divided into small areas called pixels. The size of each pixel defines the resolution of the coverage map (often between 25 to 50 meters in practice).

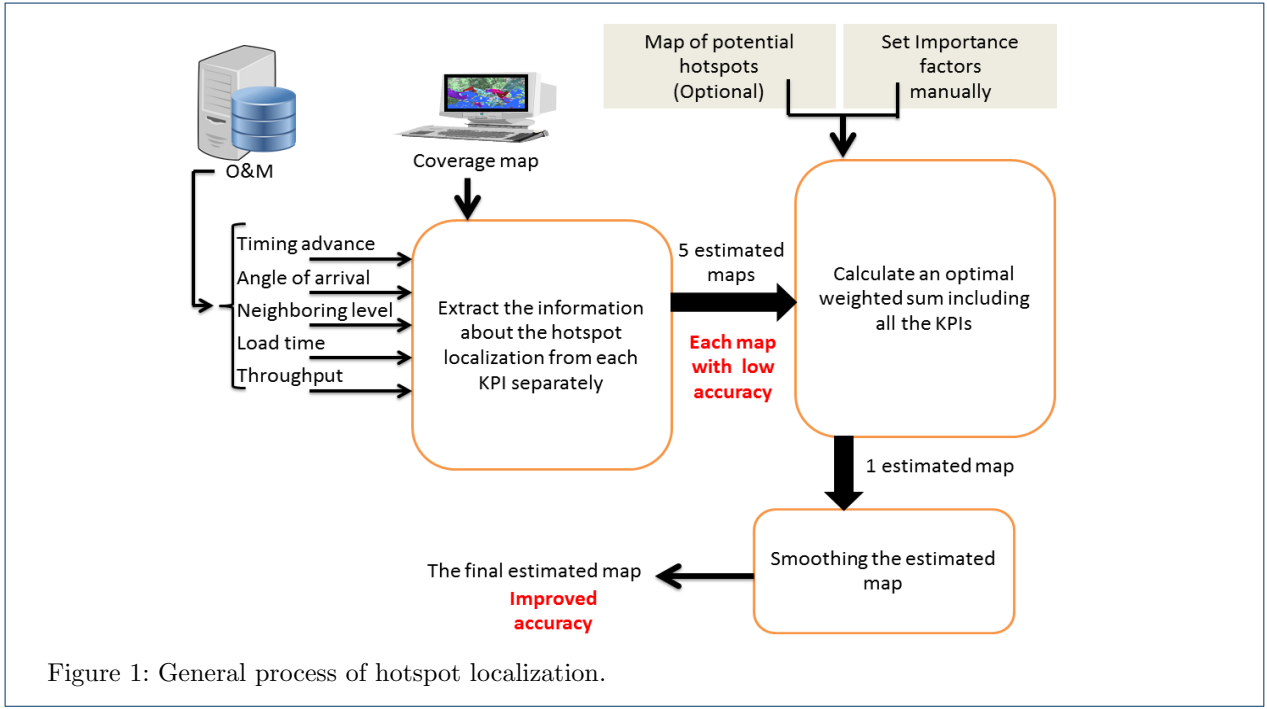

Figure 1: General process of hotspot localization.

The importance of each KPI is measured and an importance factor is assigned to it in order to avoid the over-confidence in the localization of traffic hotspots due to the correlation between KPIs. Optimal importance factors are found by solving a least square problem minimizing the distance between the estimated map and the map of potential hotspots. The map of potential hotspots represents additional information about potential hotspot zones obtained from commercial data. Then, we define a global metric combining all the five, previously-cited KPIs and using the optimal importance factors. The localization of traffic hotspot zones can give significant weights to isolated areas, or pixels, as well as pixels in the edge of a hotspot. And so, smoothing the estimated map is an additional step proposed in this paper in order to make the estimated traffic distribution more accurate.

In this paper, we are mainly interested on the combination of the O&M KPIs, projecting them over the coverage map and smoothing the estimated distribution. However, the map of potential hotspots does not play a crucial role in this method of hotspot localization. In fact, it is possible to manually tune the importance factors until having a good accuracy of localization. Whereas, a possible additional step to the localization algorithm is expected to provide better results. This step is simply projecting O&M KPIs over an available map of potential hotspots.

It is possible to add other KPIs in the algorithm of traffic localization. But, in such cases, it is essential to properly combine the KPIs in order to avoid the over-confidence due to correlated KPIs.

# 3 INPUTS FOR TRAFFIC LOCALIZATION

## 3.1 Coverage map: Definition and notation

A coverage map is a map of a bounded geographical area where the signal level from each cell is given in each pixel. The coverage map is often taken from coverage prediction tools or from real measurements such as coverage provided by the Minimization of Drive Tests (MDT) techniques [10]. In this paper, the coverage map is denoted by $\mathcal{L}$ and constitutes a bounded part of $\mathbb{R}^2$. Without loss of generality, we assume that $\mathcal{L}$ has the form of a square and is discretized into equally $m^2$ pixels.

For $1 \leq i, j \leq m$, we designate by $P_{i,j}$ the coordinates of each pixel in $\mathcal{L}$ and we assume that the pixel $P_{0,0}$ is located at the lower left corner of the map (the origin of $\mathbb{R}^2$).

We denote by $\mathcal{C} = \{C_1, C_2, ..., C_n\}$ the set of the cells located in the coverage map $\mathcal{L}$, where $C_k$ is the geographical area covered by cell $k$ and $n$ is the number of cells. It is clear that $\mathcal{C} \subseteq \mathcal{L}$ because some pixels are not covered by any cell.

Let $RSRP_{k,i,j}$ be the Reference Signal Received Power (or its equivalent the RSCP in WCDMA and $R_{xlev}$ in GSM networks) [11] from cell $k$ in pixel $P_{i,j}$,

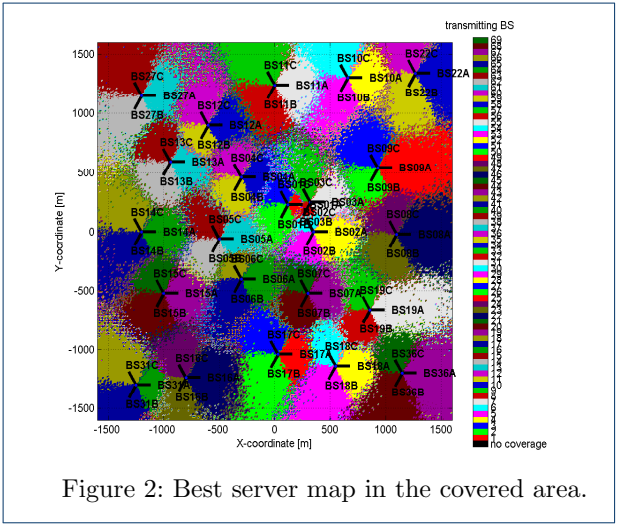

Figure 2: Best server map in the covered area.

then the cell coverage $C_k$ is given by

$$For\ 1 \leq k \leq n,\ C_k =$$
$$\{P_{i,j} \in \mathcal{L}\ such\ that\ RSRP_{k,i,j} = \max_{1 \leq l \leq n} RSRP_{l,i,j}\} \quad (1)$$

For every pixel $P_{i,j}$ in $\mathcal{L}$, we denote by $c^*_{i,j}$ and $\hat{c}_{i,j}$ respectively the index of the first and second best serving cell.

$$c^*_{i,j} = arg \max_{1 \leq l \leq n} RSRP_{l,i,j} \quad (2)$$

$$\hat{c}_{i,j} = arg \max_{\substack{1 \leq l \leq n \\ l \neq c^*_{i,j}}} RSRP_{l,i,j} \quad (3)$$

Fig. 2 illustrates an example of the coverage map with the best serving cell in each pixel. Colors are used to distinguish the area covered by each sector in the network. Each site has an identifier (BS01,BS02..) and sectors' identifier in each site is either A or B or C since we are working in a trisectorized network.

Note that in practice even if each pixel has a best serving cell, a UE located indeed in this pixel is accepted in the network only if the signal level received from its serving cell is higher than a certain threshold, called $Q_{rxlevmin}$ in 3GPP LTE standard.

### 3.2 Used Key Performance Indicators (KPIs)
*3.2.1 KPI1: Timing Advance*
Timing Advance (TA) (in GSM, LTE and LTE-A) or the propagation delay (in WCDMA) is a time offset realized by the BS between its own transmission and the transmission received from the UE. According to the calculated offset, the serving BS determines the suitable TA for the UE [12]. Then, from this TA, the BS calculates the distance traveled by the radio signal. In practice, depending on the resolution (or granularity) of TA, a specific distance range where the UE is located will be calculated. In practice, TA is used as a KPI for network supervision and analysis, with a granularity of 78.25 meters in LTE networks [12].

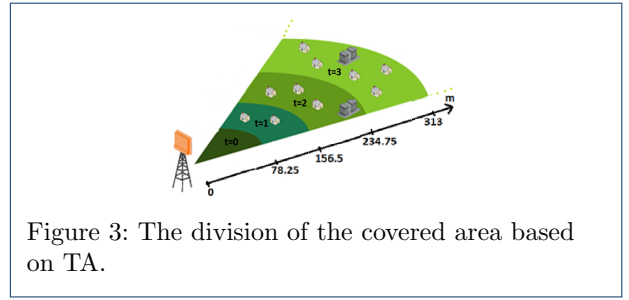

Figure 3: The division of the covered area based on TA.

Based on this granularity, TA is discretized into 6 intervals indexed by $t$ of the form $[78.25 \times t,\ 78.25 \times (t+1)]$, with $0 \leq t \leq 4$, and $[391.25,\ +\infty)$ for $t = 5$. As illustrated in Fig. 3, each cell is divided into several intervals according to the above defined ranges of distances. The TA KPI that we obtain from the O&M database is the distribution of the distance from the BS to the UEs in the cell. For example, 30% of the UEs are in the range of $t = 0$, 20% in the range of $t = 1$, 40% in the range of of $t = 2$, 10% in the range of of $t = 3$ and 0% in the range of $t = 4$ and $t = 5$.

For each cell $k$, we denote by $\tau_t(k)$ the value of the TA KPI that gives the percentage of UE in the TA interval $t$.

*3.2.2 KPI2: Angle of Arrival*
Angle of Arrival (AoA) is defined as the estimated angle of a UE with respect to a reference direction, typically the geographical North. The value of AoA is positive in an anticlockwise direction [11]. In general, any uplink signal from the UE can be used to estimate the AoA, but typically a known signal such as the Sounding Reference Signals (SRS) or DeModulation Reference Signals (DMRS) would be used [13]. The serving BS determines the direction of arrival by measuring the TA at individual elements of the antenna array and thus from these delays, the AoA is calculated.

In order to construct the shape of the downlink beam, direction of arrival estimation already exists and is supported in WCDMA but it is not standardized [14]. Experiments in [19–21] showed that the accuracy of AoA estimation depends mainly on the number of antenna elements and also on the separation distance between them. It varies from small deviations (around 2° to 5°) to significant deviations (about 30° = $\frac{\pi}{6}$) [19–21].

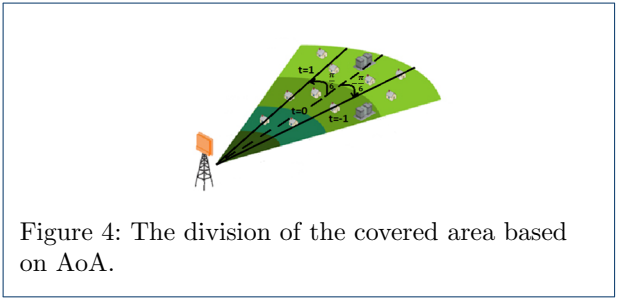

Figure 4: The division of the covered area based on AoA.

Based on the possible AoA deviations provided in [19–21], we assume that the cell is divided into 3 zones as in Fig. 4 relative to the angle between the UE, the BS and the geographical North. Likewise the TA, we denote by $t = -1, 0$ or $1$ the index of each AoA zone.

For each cell $k$, we denote by $\phi_t(k)$ the value of the AoA KPI that gives the percentage of UE in the AoA zone $t$. Therefore, the percentage of UEs connected to cell $k$ and which are in the range of $[-\frac{\pi}{6}, \frac{\pi}{6}]$ is denoted $\phi_0(k)$. Then, $\phi_1(k)$ represents the percentage of UEs connected to cell $k$ and which are in an angle of arrival larger than $\frac{\pi}{6}$. Also, $\phi_{-1}(k)$ represents the percentage of UEs connected to cell $k$ and that are in an angle of arrival less than $-\frac{\pi}{6}$. In this sectorization, we consider that the angle of arrival is equal to zero when pixel $P_{i,j}$ has the same angle as the azimuth[2] of its serving sector.

An example of distribution that we can get is 30% of UEs with range corresponding to $t = -1$, 40% for the range of AoA $t = 0$ and 30% for $t = 1$.

### 3.2.3 KPI3: Neighboring cell level

In different events of UEs connected to the network such as handover process, every UE measures the signals of the detected cells and sends a report of these measurements to its serving cell. In GSM, the detected neighboring cells list is reported periodically by the UE to the BS. However, in 3G and 4G networks, it is reported to the BS only when a special event is triggered. The counter in the O&M database representing the neighboring cell level of a given neighboring cell is incremented whenever that cell is received (in a measurement report) as an eligible candidate cell for handover (Fig. 5). In order to get the KPI as a distribution, the neighboring cell level of each neighboring cell is calculated with respect to the total number of reported neighboring cells.

For each neighboring cell $l$ of the serving cell $k$, we denote by $\vartheta_l(k)$ the relative number of times where cell

---

[2]The azimuth is the angle between the direction of maximum antenna radiation with respect to the geographical North.

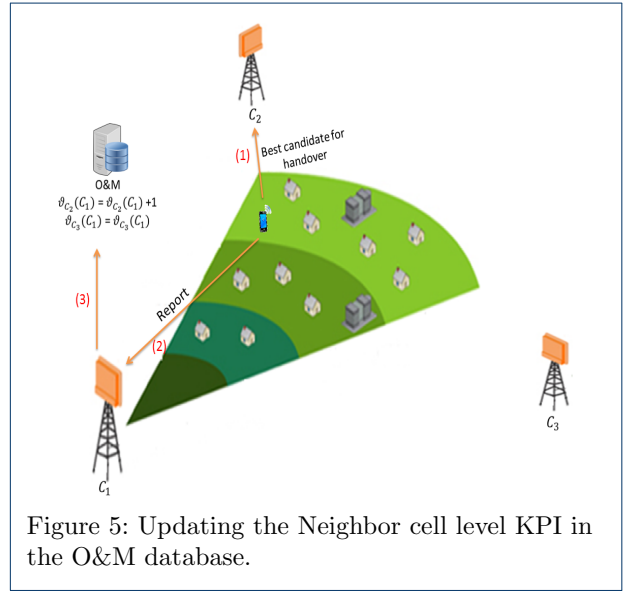

Figure 5: Updating the Neighbor cell level KPI in the O&M database.

Table 1: Neighboring cell level for the cell BS1A in Fig. 2.

| Cell ID | BS3A | BS7B | BS3C | BS6C | BS2C | BS2A | BS3B | BS4A |
|---|---|---|---|---|---|---|---|---|
| Neighboring level | 0.2478 | 0.1371 | 0.1364 | 0.1298 | 0.1038 | 0.1007 | 0.0759 | 0.0685 |

$l$ is reported as eligible candidate cell for handover in a measurement report to the serving cell $k$.

Normalizing with the total number of reported neighboring cells makes $\vartheta_l(k)$ a density distribution KPI, i.e.

$$\sum_{l \in S_k} \vartheta_l(k) = 1 \qquad (4)$$

$S_k$ is the set of the neighboring cells of the cell $k$. Table 1 shows an example of this KPI for the given serving cell BS1A (refer to Fig. 2).

### 3.2.4 KPI4: Load time

The load time represents the percentage of time when the cell's resources are fully occupied. This KPI is a gauge counter (i.e. a percentage of time) already defined and provided by all O&M vendors. It is calculated per hour and can be given for the busy hours, for the day or whether for the week or the month. In this paper, we denote by $\rho(k)$ the load time of cell $k$.

### 3.2.5 KPI5: Arithmetic and Harmonic Mean throughputs

The Arithmetic Mean Throughput (AMT) is the mean of the throughputs of all connected UEs. On the other side, the Harmonic Mean Throughput (HMT) is the average of the inverses of the throughputs[3]. The HMT differs from the AMT by the fact that it gives more importance to throughputs of UEs with bad radio conditions. Based on the formula of the cell throughput defined in the O&M database by most equipment vendors, it is important to notice that the cell throughput has almost the same value as the HMT.

The utility of these KPIs comes from the following possible scenarios: if HMT is high but always lower than AMT, one can infer that most of the users might be in bad radio conditions and if the arithmetic mean throughput is high, most of the users should be close to the serving cell. In other words, if most of the users are in the cell edge, the HMT becomes more significant and the AMT is decreased and becomes close to the HMT. However, if most of the users are in good radio conditions, the difference between the HMT and the AMT becomes significant. Moreover, in the case of uniformly distributed traffic in the cell between the edge and the center, the difference between the arithmetic and the harmonic mean throughput is also large but less than the case where UEs are mostly concentrated in the center.

For each cell $k$, we denote by $\mu_a(k)$ and $\mu_h(k)$ are respectively the AMT and the HMT KPIs taken from O&M database.

### 3.3 Map of potential traffic hotspots

A map of potential traffic hotspots is an additional layer obtained from commercial data so as to improve the hotspot localization. In this case, the spatial traffic distribution is calculated with taking into consideration the a priori knowledge of the location of industrial and commercial zones or the location of residences. This map includes also informations about the location of rivers, forests etc.. Such a map represents a reference map to define the contribution of each KPI in the estimation of the spatial traffic distribution like described in subsection 4.2.

We define $\hat{Q} = (\hat{q}_{i,j})_{1 \leq i,j \leq m}$ as the matrix representing the potential hotspots as follows

$$\hat{q}_{i,j} = \begin{cases} \omega_{i,j} & \text{if } P_{i,j} \text{ is in a potential hotspot} \\ 0 & \text{otherwise} \end{cases} \quad (5)$$

$\omega_{i,j}$ represents the importance of the potential hotspot. In fact, the weight assigned to each hotspot is evaluated according to the heaviness of the traffic that can be carried inside it. For example, a residential zone cannot take the same weight as a very large commercial zone.

---

[3]HMT is always lower than AMT according to the mathematical definition of the harmonic mean and the arithmetic mean.

## 4 TRAFFIC LOCALIZATION ALGORITHM: DESIGN AND OPTIMIZATION

### 4.1 Description of the algorithm

Before running the localization algorithm, the O&M reports the statistics of each KPI corresponding to the cells in its controlled area. Then, the following steps are performed.

*4.1.1 Step 1: calculate the spatial distribution of traffic weights according to TA.*

For the design of the step based on TA, we attribute to each pixel $P_{i,j}$ a traffic weight $q_{i,j}^{(1)}$. To do this, each cell in the coverage map is divided into 6 zones depending on the distance from the BS as illustrated in Fig. 3. Then, each pixel takes a weight equal to the percentage of UEs in its range of TA.

$$q_{i,j}^{(1)} = \sum_{t=0}^{5} \tau_t(c_{i,j}^*) \times \chi_{1,t}(P_{i,j}) \quad (6)$$

where $\chi_{1,t}(P_{i,j})$ is the indicator function of TA zone $t$ in cell $c_{i,j}^*$; $\chi_{1,t}(P_{i,j})$ takes the value 1 if the pixel belongs to TA zone $t$ of its serving cell and 0 otherwise.

*4.1.2 Step 2: calculate the spatial distribution of traffic weights according to AoA.*

In this step, each cell in the coverage map is divided into 3 zones as illustrated in Fig. 4. So, each pixel in the same range of AoA takes a traffic weight $q_{i,j}^{(2)}$ equal to the percentage of connected UEs in this range of AoA.

$$q_{i,j}^{(2)} = \sum_{t=0}^{5} \phi_t(c_{i,j}^*) \times \chi_{2,t}(P_{i,j}) \quad (7)$$

Likewise, $\chi_{2,t}(P_{i,j})$ is the indicator function of AoA zone $t$ in cell $c_{i,j}^*$; $\chi_{2,t}(P_{i,j})$ is given by

$$\chi_{2,t}(P_{i,j}) = \begin{cases} 1 & \text{if } Angle(P_{i,j}, c_{i,j}^*) - \theta(c_{i,j}^*) \in I_t \\ 0 & \text{otherwise} \end{cases} \quad (8)$$

with

$$I_t = \begin{cases} [-\frac{\pi}{6}, \frac{\pi}{6}] & \text{if t=0} \\ [\frac{\pi}{6}, \pi] & \text{if t=1} \\ [-\pi, -\frac{\pi}{6}] & \text{if t=-1} \end{cases} \quad (9)$$

and $Angle(P_{i,j}, c_{i,j}^*))$ is the angle between pixel $P_{i,j}$ and its serving cell with respect to the geographical North of the coverage map. $\theta(c_{i,j}^*)$ is the azimuth of the antenna of cell $c_{i,j}^*$.

### 4.1.3 Step 3: calculate the distribution of traffic weights according to the neighboring cell level.

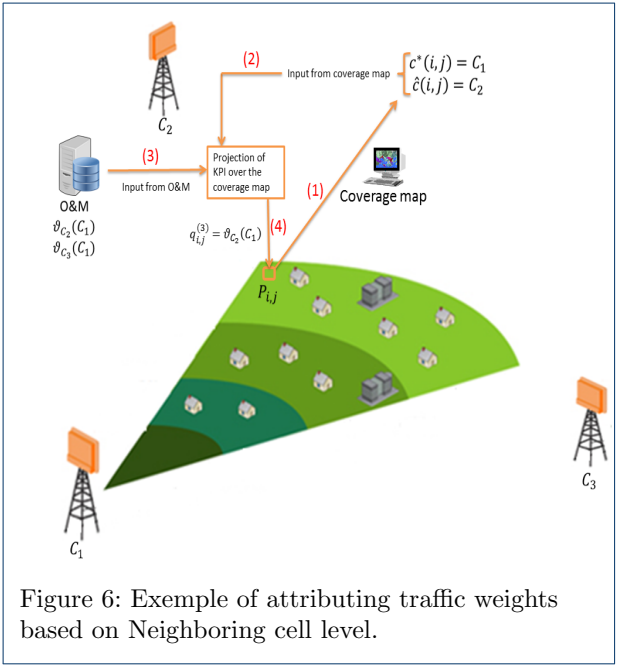

Figure 6: Exemple of attributing traffic weights based on Neighboring cell level.

Exploiting the neighboring cell level degree is motivated by the fact that when a neighboring cell is reported many times rather than others, the pixels having this cell as the second best serving cell contain probably most of the traffic.

We attribute to each pixel a traffic weight $q_{i,j}^{(3)}$ equal to the value of the neighboring cell level of its second best serving cell like shown in Fig. 6.

$$q_{i,j}^{(3)} = \vartheta_{\hat{c}_{i,j}}(c_{i,j}^*) \quad (10)$$

### 4.1.4 Step 4: calculate the distribution of traffic weights according to cell loads.

In hotspot zones, the traffic evolution has almost the same behavior in the neighboring cells due to the load balancing and forced handovers [15]. As a consequence, if a cell is congested and one of its neighbors is congested, it is likely that most of the traffic is located in the edge between the two cells and if there is no correlation with any of the neighboring cells, then traffic is most probably generated from users close to the serving cell. Furthermore, if two cells are congested but the heavy traffic is not located between them, the attributed traffic weight in the region between these two cells will be significant. The next step provides corrections to this step with comparing the arithmetic mean throughput to the arithmetic mean throughput of the cell.

We define $\psi_{c_{i,j}^*}$ as the set of neighboring cells having the same behavior in terms of load time as the serving cell $c_{i,j}^*$ and are eligible for RSRP-based handover in $P_{i,j}$. It is given by

$$\psi_{c_{i,j}^*} = \{k \in \mathcal{C} \text{ such that } |\rho(c_{i,j}^*) - \rho(k)| < \epsilon$$
$$\text{and } |RSRP_{c_{i,j}^*,i,j} - RSRP_{k,i,j}| < \lambda\} \quad (11)$$

where $\lambda$ is a parameter used to identify the set of candidate cells for a possible handover (by default, it is equal to the handover margin) and $\epsilon$ is used in order to get only the cells having the same behavior as the serving cell ($\epsilon$ is set to 0.1 in the localization exercise shown, next, in the simulation section).

The traffic weight $q_{i,j}^{(4)}$ attributed to each pixel based on the load KPI is then given as follows

$$q_{i,j}^{(4)} = \begin{cases} \frac{1}{n(\psi_{c_{i,j}^*})} \sum_{k \in \psi_{c_{i,j}^*}} \rho(k) & \text{if } \rho(c_{i,j}^*) > \tilde{\rho} \\ 0 & \text{otherwise} \end{cases} \quad (12)$$

$n(\psi)$ means the number of elements in the set $\psi$, $\tilde{\rho}$ is a threshold used to identify the very loaded serving cells (in practice the congestion threshold is set to 70%). Thus, only pixels, belonging to congested cells, get a traffic weights according to this KPI.

### 4.1.5 Step 5: calculate the distribution of traffic weights according to AMT versus HMT.

Attributed traffic weights are the difference between the AMT of the cell and the HMT divided by a constant $\mu_0$ in order to have all the calculated weights in the range [0, 1]. The constant $\mu_0$ is the maximum throughput that a user can get. The weight in each pixel depends on its position relative to the position of its best serving cell and also on the value of the difference between the two mean throughputs. So, the weights for pixels in the cell center take the value of the difference between the HMT and the AMT divided by $\mu_0$. The rest of the pixels in the cell takes a traffic weight equal to the complementary of this value as given in (13).

The traffic weight $q_{i,j}^{(5)}$ attributed to each pixel based on the HMT and the AMT is then formulated as follows:

$$q_{i,j}^{(5)} = \begin{cases} \frac{\mu_a(c_{i,j}^*) - \mu_h(c_{i,j}^*)}{\mu_0} & \text{if } RSRP_{c_{i,j}^*,i,j} > RSRP_0 \\ 1 - \frac{\mu_a(c_{i,j}^*) - \mu_h(c_{i,j}^*)}{\mu_0} & \text{if } RSRP_{c_{i,j}^*,i,j} < RSRP_0 \end{cases} \quad (13)$$

where $RSRP_0$ is a RSRP threshold used to differentiate between the cell edge versus cell center UEs.

*4.1.6 Step 6: include all KPIs in one global metric and estimate the traffic hotspot zones.*

After analyzing all the KPIs and extracting information about hotspots from them, this step identifies the way to transpose all the outputs from the previous steps into a unique traffic weight $Q = (q_{i,j})_{1 \leq i,j \leq m}$. The simplest way to get the matrix $Q$ is by making a weighted sum of the matrices $Q^{(s)} = (q_{i,j}^{(s)})_{1 \leq i,j \leq m}$, $1 \leq s \leq 5$, with an importance factor $\mathbf{x} = (x_1, x_2, x_3, x_4, x_5)^T$ depending on the importance of the information extracted from each KPI. So, we have that

$$Q = \sum_{s=1}^{5} x_s Q^{(s)} \qquad (14)$$

The trivial value of $\mathbf{x}$ is $(\frac{1}{5}, \frac{1}{5}, \frac{1}{5}, \frac{1}{5}, \frac{1}{5})^T$. In this case, $\mathbf{x}$ is not optimal because some KPIs provides more precision than others and some KPIs are correlated, as said before. For these reasons, we define the optimal value of $\mathbf{x}$ as the result of an optimization problem which reduces the distance between the map of potential hotspots and the matrix $Q$. This optimization is detailed in next subsection 4.2.

*4.1.7 Step 7: smoothing of the estimated spatial distribution of the traffic.*

The estimated spatial traffic distribution is further enhanced with smoothing which consists in combining the values estimated in each pixel with the values estimated in the neighboring pixels. The main advantage of smoothing [16] is to delete isolated pixels with significant weights (wrong estimated hotspots). From another side, this step allows to ensure more the existing hotspots in the network once they are not eliminated.

It was shown in [17] that the spatial traffic distribution follows a Lognormal distribution or a mixture of Lognormal distributions. Smoothing the estimation with a Lognormal smoother supposes the apriori knowledge of the direction of its shape relative to the angle coordinate. Therefore, we choose the function of the smoother to be a Gaussian distance decay (symmetric with respect to the angle coordinate).

$$q_{i,j}^* = \frac{\sum_{P_{i',j'} \in \mathcal{L}} G(P_{i,j}, P_{i',j'}) q_{i',j'}}{\sum_{P_{i',j'} \in \mathcal{L}} G(P_{i,j}, P_{i',j'})} \qquad (15)$$

where

$$G(P_{i,j}, P_{i',j'}) = \frac{1}{\sqrt{2\pi h}} e^{-\frac{|P_{i,j} - P_{i',j'}|^2}{2h}} \qquad (16)$$

with $|P_{i,j} - P_{i',j'}|$ is the euclidean distance between the pixels $P_{i,j}$ and $P_{i',j'}$ and $h$ is a parameter related to the number of the neighboring pixels that are involved into the smoothing of the value in pixel $P_{i,j}$ (in simulations, we take $h = 10^{-3} m^2$). $q_{i,j}$ are the elements of the weight matrix obtained from the previous step and $q_{i,j}^*$ are the elements of the weight matrix estimated after smoothing.

### 4.2 Optimization of the localization algorithm

It is possible to manually set the importance factors but the relative performance of the localization process may be incompetent comparing to existing solutions. For instance, KPIs holding small information should not be well considered in the aggregated traffic weight $Q$ because they bias the results. In order to improve the accuracy of the hotspot localization algorithm, the importance factors, as defined in step 6, need to be optimized so that the contribution of each KPI in the weighted sum is justified.

Finding the optimal vector $\mathbf{x}$ is the result of an optimization problem that reduces the distance between the available map of the potential hotspots $\hat{Q}$ and the estimated traffic map $Q$ found in Step 6 of the localization algorithm. The optimization is performed only one time for an area that we well know its potential hotspots. The obtained optimal importance factor might be used later for any similar environment (urban, suburban, rural...) without any knowledge of its potential hotspots because the contribution of each KPI mostly depends on the type of environment. For instance, TA and AoA are more efficient and have more contribution in rural areas, however in urban areas, measuring the TA or the AoA presents more errors due to obstacles resulting thus to more reflexions and diffractions.

The optimization of $\mathbf{x}$ can be formulated as follows:

$$\begin{cases} \underset{x \in \mathbb{R}^5}{minimize} \sum_{i,j=1}^{m} (q_{i,j} - \hat{q}_{i,j})^2 = \\ \qquad \qquad \sum_{i,j=1}^{m} \left( \sum_{s=1}^{5} x_s q_{i,j}^{(s)} - \hat{q}_{i,j} \right)^2 \\ x_s \geq 0, \ s = 1..5 \end{cases} \qquad (17)$$

In order to put the optimization problem in a known and solvable form, we consider

$$A^T = \begin{bmatrix} q_{1,1}^{(1)} & \cdots & q_{1,m}^{(1)} & \cdots & q_{2,1}^{(1)} & \cdots & q_{m,m}^{(1)} \\ \cdot & \cdot\cdot & \cdot & \cdot\cdot & \cdot & \cdot\cdot & \cdot \\ q_{1,1}^{(5)} & \cdots & q_{1,m}^{(5)} & \cdots & q_{2,1}^{(5)} & \cdots & q_{m,m}^{(5)} \end{bmatrix}$$

$$\mathbf{b} = [\hat{q}_{1,1}...\hat{q}_{1,m}...\hat{q}_{2,1}...\hat{q}_{m,m}]^T$$

Clearly, $A$ is a matrix of $m^2$ lines and 5 columns and $\mathbf{b}$ is a vector of length $m^2$.

Using matrix notation, the optimization problem of equation (17) becomes a minimization of the distance between the vectors $A\mathbf{x}$ and $\mathbf{b}$ in the space $\mathbb{R}^{m^2}$:

$$\begin{cases} \underset{\mathbf{x}\in\mathbb{R}^5}{minimize} \|A\mathbf{x}-\mathbf{b}\|_2 \\ x_s \geq 0 \, , \, 1 \leq s \leq 5 \end{cases} \quad (18)$$

where $\|.\|_2$ is the standard 2-norm (the Euclidian norm) in the space $\mathbb{R}^{m^2}$.

This formulation represents a least square optimization problem and can be solved using the Gauss Newton method [18]. This method is fast and provides accurate solution. As said before, the optimal importance factors of this problem is not specific to limited number of possible scenarios. It can indeed be changed from a scenario to another but with small variations since the importance of each KPI remains more or less the same and is slightly independent from the taken scenarios. The optimal vector $\mathbf{x}$ would lead to a more precise hotspot localization that we look forward to in the objective of the paper.

## 5 NUMERICAL RESULTS

### 5.1 Parameters' settings

In order to validate and evaluate the proposed algorithm, we use a LTE simulator that allows dynamic users' arrivals and departures after being served. At each time step of the simulator, equal to $1s$, the number of call attempts are generated according to a Poisson distribution with intensity $\lambda = 200$ UEs/s. UE positions are generated depending on the traffic weight in each pixel of the coverage map. UEs are accepted in the network only when there are available resources and when their RSRP is higher than the threshold $Q_{rxlevmin}$ (set to $-115dBm$ in this work). We suppose that each UE has a file of size $1Mbit$ to download. Each UE quits the network after downloading the entire file. Moreover, we consider that a part of UEs (20%) are moving during their transmission with a mobility of $8.33km/h$. This means that the simulation supports also handover events with a margin set to $6dB$. UEs are scheduled according to the round robin model. At the end of the simulation, which lasts 1 hour, all KPIs (including the previously cited KPIs) are calculated and stored in a file.

The simulated network is shown in Fig. 2 and composed of 23 tri-sectorized eNB (evolved Node-B) covering an area of $3\times 3Km^2$. eNB positions and physical characteristics are taken from a real network. Each sector in each eNB is equipped with a directive antenna having a beamwidth equal to $65^o$. Moreover, each sector has an available capacity equivalent to $20Mhz$ of spectral bandwidth operating in the band $2.6Ghz$. Because of the network heterogeneity, the list of parameters (e.g., eNB transmitter noise figure, cable loss, maximum transmit power...), involved in this simulation depends on each site. Hence, it is worthless to cite all these parameters.

The coverage map (RSRP from each cell in each pixel), provided by a planning tool, has a resolution equal to 25 meters. According to this map, the first and the second best serving cell are identified in each pixel.

Before starting the simulation, we attribute the traffic weight in each pixel of the coverage map based on a mixture of Log-normal distributions [17]. As shown in Fig.7, each peak of a Log-normal density represents a traffic hotspot.

For the potential traffic map, we attribute arbitrary weights in some pixels according to our knowledge about some hotspot zones in the chosen map. The exploitation of this map improves the precision of hotspot localization but recall again that it is not a key element in the proposed algorithm.

The validation of the algorithm consists in finding the traffic weight in each pixel (following the steps of the algorithm) and comparing it with the original traffic weights generated at the beginning of the simulation.

### 5.2 Results

Running the hotspot localization algorithm and performing the optimization, as described in Subsection 4.2, we get the importance factor

$$\mathbf{x} = [0.418 \ \ 0.2689 \ \ 0.2281 \ \ 0.0358 \ \ 0.0491]^T$$

The validation of the obtained importance factors is not the objective of the present paper. However, we were able to test other values set for importance factors and the relative results are clearly improved after the optimization step. So, these optimal importance factors are used in the performance evaluation for the different reasons cited before.

In Fig. 7, the original distribution of the traffic generated at the beginning of the simulation is drawn. In Fig. 8a, the estimated traffic distribution is depicted based on the proposed algorithm without including the smoothing process. Fig. 8b represents the spatial distribution of the traffic after smoothing the calculated distribution.

Regarding the identification of the hotspot zones, we observe that most of the hotspot zones are found in the entire network map.

In Table 2, the coordinates of the generated hotspots and those estimated in step 6 and in step 7 are extracted from the calculated matrices generating the

Table 2: Coordinates of generated and estimated hotspots holding highest traffic weights (x,y) in meters.

| Generated hotspot | (1100,960) | (760,940) | (20,980) | (-1020,-640) | (1040,280) | (-200,-120) | (-960,740) | (220,100) | (960,-300) |
|---|---|---|---|---|---|---|---|---|---|
| Estimated hotspot in step 6 | (1140,940) | (740,940) | (40,980) | (-1040,-660) | (1200,420) | (-240,-180) | (-940,780) | (320,140) | (940,-280) |
| Estimated hotspot in step 7 | (1100,960) | (760,900) | (40,980) | (-1020,-660) | (1100,340) | (-180,-120) | (-920,760) | (260,100) | (960,-320) |
| Distance between the original and the estimated hotspot in step 6 | 44.72 | 20 | 20 | 28.28 | 212.6 | 72.11 | 44.72 | 107.7 | 28.28 |
| Distance between the original and the estimated hotspot in step 7 | 0 | 40 | 20 | 20 | 84.85 | 20 | 44.72 | 40 | 20 |

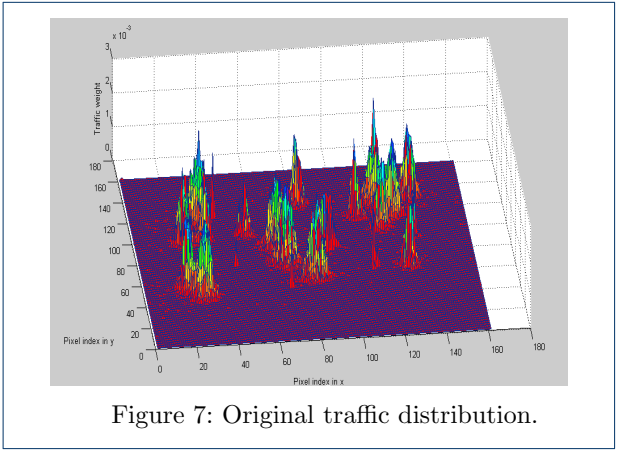

Figure 7: Original traffic distribution.

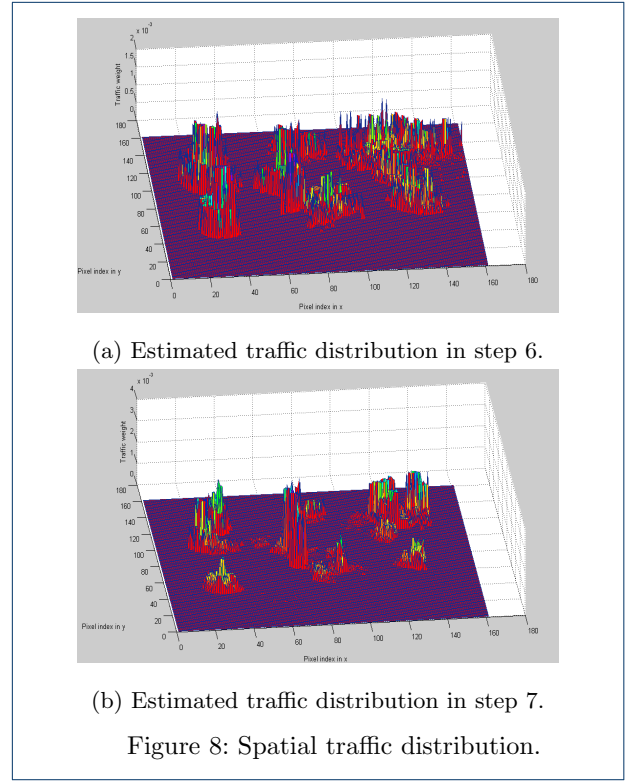

(a) Estimated traffic distribution in step 6.

(b) Estimated traffic distribution in step 7.

Figure 8: Spatial traffic distribution.

figures above. From Table 2, we find out that the execution of the proposed algorithm until the step 6 pinpoints the hotspots generated in the network with a precision of 59.84 meters (see Table 2, the mean of the calculated distances between the original and the estimated hotspots in step 6). However, the spatial traffic distribution is quite uniform inside these hotspots. Then, the shape of the estimated traffic distribution is improved after smoothing and the precision becomes about 31 meters (see Table 2, the mean of the calculated distances between the original and the estimated hotspots in step 7).

Fig. 9 shows the CDF of traffic weights attributed to pixels based on the proposed algorithm. To localize the hotspot zones, we are interested only in the region surrounded with blue ellipse in Fig. 9 which represent the significant weights.

Fig. 9 indicates that using all the 5 KPIs in the network gives better estimation of traffic distribution as compared to the case when using only some of them. In fact, significant weights have a small density when using only one KPI which means that the traffic distribution is more flat and is uniform inside the hotspot zone. This density is increased and becomes near the exact distribution when all the KPIs are used and when all the steps of the proposed algorithm are performed.

From Fig. 9, we note that the most useful KPIs are the TA, AoA and the neighbor cell level. The mean throughput and the load time have also an impact on the estimation and improve further the localization.

In addition, smoothing the estimated traffic distribution reduces the difference between the original spatial distribution of the traffic and the estimated one.

Among the plotted curves in Fig. 9, the curve representing the combination between TA and neighboring cell level allows us to compare approximately the solution proposed in [5] to the present proposed algorithm. So, the new proposed algorithm perform better and provides promising results comparing to the solution disclosed in [5].

We compare in Table 3 the percentage of detected hotspots that have the highest weights to the original highest hotspots in the network.

In Table 3, the analysis is based on the fact that weights are sorted in a decreasing way. We determine the coordinates of the pixels which hold 0.5% (1, 2, 5, 10, 20, and 50 respectively) of the traffic in the network (using the real distribution). Then, we calculate the sum of estimated weights in these pixels.

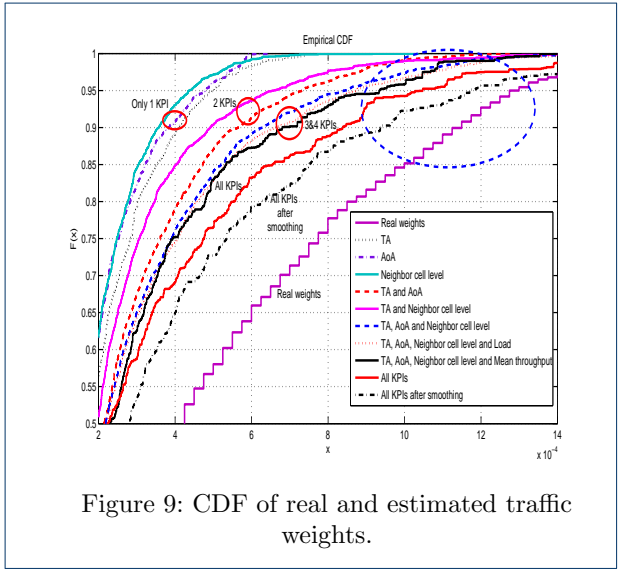

Figure 9: CDF of real and estimated traffic weights.

Table 3: Percentage of detected hotspots.

| Real % of first hotspots | Estimated % with only TA | Estimated % with all KPIs | Estimated % after smoothing |
|---|---|---|---|
| 0.5 | 0.12 | 0.30 | 0.422 |
| 1 | 0.26 | 0.48 | 0.58 |
| 2 | 0.46 | 1.09 | 1.13 |
| 5 | 1.12 | 2.17 | 2.56 |
| 10 | 2.29 | 4.34 | 4.8 |
| 20 | 4.87 | 33.45 | 9.7 |
| 50 | 14.22 | 74.34 | 27.32 |

The purpose of the evaluation in Table 3 is to see the importance of using all the KPIs and the impact of smoothing on the localization of most significant hotspots. From another side, the column related to performances of using only TA stands to the reason that, in literature, several traffic localization techniques are based on TA. So, this allows a common comparison of these techniques to the proposed one. In this context, Table 3 shows that the hotspot localization is clearly improved when it is based on the 5 KPIs comparing to the use of only TA.

From Table 3, we also observe that the first hotspots that have the highest weights are reasonably estimated and this estimation is further enhanced with the step of smoothing. However, when we increase the percentage of the first significant hotspots (from 0.5% to 10%), the estimation shows less efficiency and this is due to the fact that some zones close to the hotspots could take a significant weight without carrying heavy traffic.

Before smoothing, pixels within the same region of a hotspot take the same weight. However, weights in the center of a hotspot are increased after smoothing and weights of pixels in the edge of a hotspot are reduced. Hence, when the percentage of the first significant hotspots is low, only pixels in the center of a hotspot are evaluated and the sum of weights before smoothing is less than the sum after smoothing. In contrast, when the percentage of the first significant hotspots is high, pixels in the edge of the hotspots are also included in the evaluation. As a result, the sum of weights before smoothing is larger than the sum after smoothing since weights of pixels in the edge of hotspots are more reduced after smoothing.

# 6 Conclusion

We have proposed in this work a modular and optimized algorithm that consists in combining several KPIs along with coverage and potential traffic hotspots map. Results showed acceptable localization error, in cases of moderate and heavy traffic. Therefore, the use of O&M KPIs projected over a coverage map can be efficient for hotspot localization as they yield promising results at low operational costs. This method is a good solution that can be used both to identify areas where a small cell must be deployed and to perform appropriate configurations to lessen the congestion rate in hotspot zones.

In the future works, we intend to study the impact of imperfect hotspot localization on the network performance. This can be done by analyzing the extra interference that would result from a bad positioning of the small cell and the degraded throughput experienced by the involved UEs in the network.

**Author details**
[1]Orange Labs, 38/40 avenue General Leclerc, 92794 Issy-les-Moulineaux, France. [2]Telecom SudParis, 9 street Charles Fourier, 91011 Evry, France.